\documentclass[twocolumn,showpacs,showkeys,amsmath,amssymb]{revtex4}
\usepackage{graphicx} 
\usepackage{color}
\usepackage{ulem} 
\newcommand{\lasrco}{La$_{1-x}$Sr$_{x}$CoO$_3$}

\newcommand{\prcapet}{Pr$_{0.5}$Ca$_{0.5}$CoO$_3$}

\newcommand{\lasrtri}{La$_{0.7}$Sr$_{0.3}$CoO$_3$}

\newcommand{\prcatri}{Pr$_{0.7}$Ca$_{0.3}$CoO$_3$}

\newcommand{\ndcatri}{Nd$_{0.7}$Ca$_{0.3}$CoO$_3$}
\newcommand{\prycatri}{(Pr$_{1-y}$Y$_{y}$)$_{0.7}$Ca$_{0.3}$CoO$_3$}
\newcommand{\laco}{LaCoO$_3$}

\newcommand{\prco}{PrCoO$_3$}

\newcommand{\st}{$\rm ^o$}

\newcommand{\emumoloe}{emu~mol$^{-1}$Oe$^{-1}$}
\newcommand{\figwp}{0.95}
\begin{document}
\sloppy
\title{The spin-state crossover and low-temperature magnetic state in yttrium doped Pr$_{0.7}$Ca$_{0.3}$CoO$_3$}
%
\author{K. Kn\'{\i}\v{z}ek}
\author{J. Hejtm\'{a}nek}
\author{M. Mary\v{s}ko}
\author{P. Nov\'{a}k}
\author{E. \v{S}antav\'{a}}
\author{Z. Jir\'{a}k}
\affiliation{Institute of Physics ASCR, Cukrovarnick\'a 10, 162 00 Prague 6, Czech Republic.}
\author{T. Naito}
\author{H. Fujishiro}
\affiliation{Faculty of Engineering, Iwate University, 4-3-5 Ueda, Morioka 020-8551, Japan.}
\author{C. R. dela Cruz}
\affiliation{Neutron Scattering Science Division, ORNL, Oak Ridge, Tennessee 37831, United States}
\begin{abstract}
The structural and magnetic properties of two mixed-valence cobaltites with formal population of
0.30 Co$^{4+}$ ions per f.u., (Pr$_{1-y}$Y$_{y}$)$_{0.7}$Ca$_{0.3}$CoO$_3$ ($y=0$ and 0.15), have been studied down to very low
temperatures by means of the high-resolution neutron diffraction, SQUID magnetometry and heat
capacity measurements. The results are interpreted within the scenario of the spin-state
crossover from a room-temperature mixture of the intermediate spin Co$^{3+}$ and low spin
Co$^{4+}$ (IS/LS) at the to the LS/LS mixture in the sample ground states. In contrast to the
yttrium free $y=0$ that retains the metallic-like character and exhibits ferromagnetic ordering
below 55~K, the doped system $y=0.15$ undergoes a first-order metal-insulator transition at
132~K, during which not only the crossover to low spin states but also a partial electron
transfer from Pr$^{3+}$ 4f to cobalt 3d states take place simultaneously. Taking into account
the non-magnetic character of LS Co$^{3+}$, such valence shift electronic transition causes a
magnetic dilution, formally to 0.12 LS Co$^{4+}$ or 0.12 $t_{2g}$ hole spins per f.u., which is
the reason for an insulating, highly non-uniform magnetic ground state without long-range order.
Nevertheless, even in that case there exists a relatively strong molecular field distributed
over all the crystal lattice. It is argued that the spontaneous FM order in $y=0$ and the
existence of strong FM correlations in $y=0.15$ apparently contradict the single $t_{2g}$ band
character of LS/LS phase. The explanation we suggest relies on a model of the defect induced,
itinerant hole mediated magnetism, where the defects are identified with the magnetic high-spin
Co$^{3+}$ species stabilized near oxygen vacancies.
\end{abstract}
\maketitle

\section{Introduction}

Perovskite cobaltites display a wide variety of structural and physical properties in dependence
on the composition and temperature. Two distinct behaviors can be identified. One is
characteristic for the undoped \laco\ and its rare-earth analogs. Their insulating ground state
derives from Co$^{3+}$ ions in the diamagnetic low-spin (LS) states. With increasing temperature
two spin-state crossovers take place. First, the paramagnetic high-spin (HS) states are induced by
thermal excitation and are gradually stabilized, which results a spin-state disproportionated
phase with strong HS/LS nearest neighbor correlations or even short-range orderings
\cite{RefGoodenough1958JPCS6_287}, that can be classified as Mott insulator. At elevated
temperature the correlations melt \cite{RefBari1972PRB5_4466}, and a more uniform phase of
quasi-metallic character is established. To account for this change and to explain the
paramagnetic properties actually observed, the high-temperature phase of \laco\ was tentatively
described as consisting of the intermediate-spin (IS) of Co$^{3+}$ species ($S=1$,
$m_{eff}=2.83~\mu_B$)
\cite{RefKyomen2005PRB71_024418,RefJirak2008PRB78_014432,RefKnizek2009PRB79_014430GGA}. It should
be noted, however, that DMFT calculations suggest for the high-temperature phase of \laco\ a
complex global state with the main weight of LS and HS states with only short visits to IS
configurations. \cite{RefKrapek2012PRB86_195104}.

Another extreme case are the metallic ground states with bulk ferromagnetic (FM) ordering, known
for hole-doped systems \lasrco\ above a critical concentration $x_c=0.22$
\cite{RefCaciuffo1999PRB59_1068,RefWu2003PRB67_174408}. Recent DMFT calculations suggest also for
this case a complex distribution of local cobalt valences and spin states
\cite{RefAugustinsky2013PRL110_267204}, but with certain simplification, the metallic nature of
these FM phases can be related to electron transfer between neighbors of the IS Co$^{3+}$/LS
Co$^{4+}$ or HS Co$^{3+}$/HS Co$^{4+}$ kinds, eventually also LS Co$^{3+}$/LS Co$^{4+}$
\cite{RefSboychakov2009PRB80_024423}. At least in the compositional region close above $x_c$, the
\lasrco\ systems seem to be dominated by a dynamic mixture of IS Co$^{3+}$ and LS Co$^{4+}$ and is
described for illustration as the IS/LS phase. This conclusion finds a support among others by the
observed values of ordered ferromagnetic moments, making in particular $1.70\mu_B$ per f.u. for
$x=0.30$.

When large cations in \lasrtri\ are substituted by rare-earth or calcium ions of smaller size, the
FM ordering still exists but the spontaneous moments actually observed are much suppressed.
Although this effect was originally related to a phase separation into FM and non-FM, the final
state at the lowest temperature is now proved, based mainly on the uniformity of molecular field
acting on Nd$^{3+}$ moments in \ndcatri, \cite{RefJirak2013JPCM25_216006}, to be essentially of
single FM phase. The observed spontaneous moments, approaching in \ndcatri\ and \prcatri\ a
limiting low value of 0.30~$\mu_B$ per f.u., can be thus ascribed to a stabilization of the LS/LS
phase, i. e. to the alternative ground state of the hole-doped cobaltites, which consists of a
dynamic LS Co$^{3+}$/LS Co$^{4+}$ mixture.

The present paper deals namely with \prcatri\ and related compounds \prycatri. These systems, all
with 30\% doping, show a quasi-metallic conductivity at ambient conditions, but in contrast to the
formation of FM state in pure \prcatri, the samples with partial substitution of Pr$^{3+}$ by
isovalent Y$^{3+}$ (for $y>0.06$) exhibit a first order transition to a weakly paramagnetic
insulating state.  Let us note that this distinct transition was observed for the first time in
the 50\% doped cobaltite \prcapet\ upon a cooling below a critical point $T_{MI}\sim 80$~K, and
the change of electric properties was accompanied with important volume, magnetic and heat
capacity anomalies \cite{RefTsubouchi2002PRB66_052418,RefTsubouchi2004PRB69_144406}. Later on,
similar transition and anomalies were found also in other praseodymium based systems in larger
region of doping. A question has been raised why sharp transition is encountered solely in
praseodymium containing cobaltites. As suggested by GGA calculations and some new experimental
results, including analysis of the temperature change of interatomic lengths and XANES spectra in
\prcapet, the reason for stabilization of the insulating low-temperature phase is a shift of the
mixed valence Co$^{3+}$/Co$^{4+}$ toward pure Co$^{3+}$, enabled by valence change of some
Pr$^{3+}$ ions to Pr$^{4+}$ ones. The decisive factor is, therefore, the exceptional closeness in
energy of the two praseodymium states. The valence shift upon phase transition has been further
confirmed and determined quantitatively by observation of the Pr$^{4+}$ related Schottky peak in
heat capacity measurements \cite{RefHejtmanek2013EPJB86_305} and by X-ray absorption spectroscopy
at Pr $L_3$ edge \cite{RefGarciaMunoz2011PRB84_045104,RefFujishiro2012JPSJ81_064709}. In addition
to the valence shift, a crossover from the IS or mixed LS/HS Co$^{3+}$ state to the LS states has
been evidenced by drop of the paramagnetic susceptibility and by low values of effective moments
below $T_{MI}$, as well as by a detailed fit of the Co X-ray absorption and emission spectra
\cite{RefHerreroMartin2012PRB86_125106}.

\begin{figure}
\includegraphics[width=\figwp\columnwidth,viewport=0 375 545 827,clip]{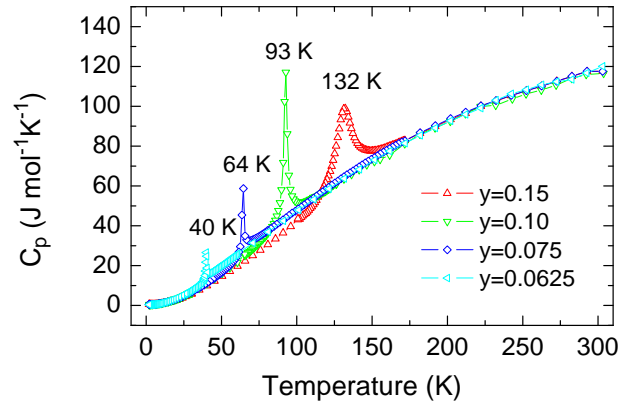}
\caption{Temperature dependence of the specific heat of \prycatri\ ($y = 0.0625-0.15$).
The critical temperatures of $M-I$ transitions are marked.}
\label{FigHC}
\end{figure}

In contrast to the prototypical compound \prcapet, which is difficult to prepare because of
problems with oxygen deficiency and phase separation, the less doped systems \prycatri\ appeared
more suitable for experimental studies. The presence of $M-I$ transition is demonstrated
indirectly by peaks in heat capacity at $T_{MI}=40$, 64~K, 93~K and 132~K for $y$=0.0625, 0.075,
0.10 and 0.15, respectively (see Fig.~\ref{FigHC}). The stabilization of tetravalent praseodymium
in the low-tempearture phase has been indicated by appearence of the low-temperature Schottky
peaks, arising due to Kramers degeneracy of Pr$^{4+}$ states and Zeeman splitting of ground
doublet by action of the molecular and external magnetic fields. The quantitative analysis
determines that the Pr$^{4+}$ population varies between 0.11 and to 0.18 per f.u., which means
that the hole concentration is decreased upon the transition from the original 30\% level to 19\%
doping in $y$=0.0625 and 12\% doping in $y$=0.15
\cite{RefHejtmanek2010PRB82_165107,RefHejtmanek2013EPJB86_305}. (No valence change was observed in
pure \prcatri.) The motivation for present study is to elucidate the character of the Pr$^{4+}$
species formed below the $M-I$ transition and to investigate the microscopic origin of
unexpectedly strong molecular field of about 17~kOe, which acts on Pr$^{4+}$ pseudospins in the
low-temperature insulating phase. The study includes neutron diffraction and magnetic measurements
on two selected \prycatri\ samples ($y=0$ and 0.15) down to 0.25~K.

\section{Experiments and calculations}

\prycatri\ ($y$=0 and 0.15) samples were prepared by a solid-state reaction as described
elsewhere~\cite{RefHejtmanek2010PRB82_165107}. The powder neutron diffraction was performed on
diffractometer Hb2a at Oak Ridge National Laboratory. The scans were recorded at selected
temperatures between the room temperature and 0.25~K. Two crystal monochromators (Ge113 and Ge115)
were used, providing neutron wavelengths $\lambda=2.408$~\AA\ and 1.537~\AA, respectively. Data
were collected between 8\st\ and 126\st\ of 2$\theta$ with the step of 0.08\st. Structural
refinements were done by Rietveld profile analysis using program FULLPROF (Version 5.30 -
Mar2012-ILL JRC).

The magnetic properties were investigated in the temperature range $2-400$~K using Quantum Design
DC superconducting quantum interference device (SQUID) MPMS XL magnetometer. For the DC extraction
magnetization measurements up to 140~kOe the QD PPMS AC/DC magnetometer with the ACMS Option was
used. The experiments included the field-cooled (FC) and zero-field-cooled (ZFC) susceptibility
scan in range $2-300$~K using DC field of 20 Oe, together with the low-temperature measurements of
virgin magnetization curves and complete hysteresis loops. In addition, the exchange bias
experiments at 2~K were performed on field-cooled samples for bias fields from 500~Oe to 70~kOe.
The thermoremanence was measured after cooling the sample in high field (typically 70~kOe) down to
2~K, switching off the field and scanning the magnetization at zero field on sample warming.

In order to determine the ground state properties of \prycatri\ cobaltites, the low-temperature
magnetic contribution of the Pr$^{3+}$ and Pr$^{4+}$ ions was theoretically analyzed. For this
sake the crystal field splitted 4$f^n$ electronic levels were calculated using 'lanthanide'
package \cite{RefEdvardsson2001CPC133_396}. This program determines takes into account the
free-ion (atomic) and crystal field terms, as well as interactions with the molecular and external
magnetic fields. As to the spherically symmetrical free-ion Hamiltonian is concerned, it depends
on many parameters, the values of which are either known experimentally or may be calculated - for
details see e.g. Ref. \cite{RefHufner1978}. The crystal field Hamiltonian represents more
formidable problem. In the present case of C$_s$ symmetry of rare-earth sites, the single-electron
crystal field is described by fifteen parameters. The best choice for Pr$^{3+}$ in \prycatri\ is
the set of crystal field parameters determined very recently for \prco\ using a first principle
method \cite{RefNovak2013JPCM}. The output of the 'lanthanide' calculation shows that the
$^3$H$_4$ multiplet of Pr$^{3+}$ (4$f^2$) splits into nine non-magnetic singlets and the ground
state properties are manifested by anisotropic Van Vleck susceptibility with components
$\chi_a=0.0254$~\emumoloe, $\chi_b=0.0181$~\emumoloe\ and $\chi_c=0.0123$~\emumoloe along the
orthorhombic axes of P$bnm$ structure. This yields an average value $\chi_{vV}=0.018$~\emumoloe
for the polycrystal, which is in excellent agreement with experimental data on a related \prco\
system in Ref. \cite{RefNovak2013JPCM}, namely with its temperature nearly independent
susceptibility in the range up to $T\sim40$~K, i.e. before the first excited level at 11~meV
starts to be populated.

The calculation for Pr$^{4+}$ (formally 4$f^1$ configuration but with large 4$f^2\underline{L}$
admixture) is subjected to more uncertainty and, moreover, there are no orthoperovskites with
Pr$^{4+}$ majority, which would allow an experimental check. As a first estimate, the values of
crystal field parameters can taken from very detailed optical spectroscopic study of TbAlO$_3$
\cite{RefGruber2008JLUM128_1271}, taking into account that this aluminate possesses the same
$Pbnm$ orthoperovskite structure and practically identical octahedral tilting as the present
\prycatri\ ($y=0.15$) sample in the low-temperature phase. The calculation shows that the
$^2F_{5/2}$ multiplet of Pr$^{4+}$ splits into three energy distant Kramers doublets (see Appendix
of Ref. \cite{RefJirak2013JPCM25_216006}. The magnetic moments associated with ground doublet
($J'=1/2$) are given by anisotropic $g$-factor with principal components $g_x=3.757$,  $g_y=0.935$
and $g_z=0.606$. (Here, the local axes $x$ and $y$ are turned with respect to the main axes of the
$Pbnm$ structure, making for two inequivalent rare-earth sites a rotation to $\pm 36$\st\ around
the orthorhombic axis $c$).

\section{Results}

\subsection{Neutron diffraction study.}

\begin{table*}
\caption{Structural parameters for \prycatri\ with $y=0$ within the $Pbnm$ space group. Refinable
atom coordinates: PrCa $4c$(x,y,1/4), Co $4b$(1/2,0,0), O1 $4c$(x,y,1/4), O2 $8d$(x,y,z)}
\begin{tabular*}{0.65\textwidth}{@{\extracolsep{\fill}} l|rrrr}
\hline
T (K)  & 0.25      & 2         & 40         & 298         \\
\hline
a (\AA)& 5.3468(2) & 5.3465(2) & 5.3463(4)  & 5.3637(2)   \\
b (\AA)& 5.3359(2) & 5.3353(2) & 5.3352(4)  & 5.3516(2)   \\
c (\AA)& 7.5386(3) & 7.5383(2) & 7.5388(5)  & 7.5702(4)   \\
\hline
x,Pr   & -0.0048(9)& -0.0052(8)& -0.0047(11)& -0.0066(10) \\
y,Pr   & 0.0347(7) & 0.0323(5) & 0.0350(8)  & 0.0277(4)   \\
x,O1   & 0.0687(7) & 0.0676(5) & 0.0704(8)  & 0.0637(8)   \\
y,O1   & 0.4921(8) & 0.4912(5) & 0.4923(9)  & 0.4901(11)  \\
x,O2   & -0.2859(7)& -0.2838(4)& -0.2880(8) & -0.2807(9)  \\
y,O2   & 0.2818(9) & 0.2855(4) & 0.2820(10) & 0.2807(9)   \\
z,O2   & 0.0356(3) & 0.0358(2) & 0.0356(6)  & 0.0319(8)   \\
\hline
\end{tabular*}
\label{tab00}
\end{table*}

\begin{table*}
\caption{Structural parameters for \prycatri\ with $y=0.15$ within the $Pbnm$ space group.
Refinable atom coordinates: PrYCa $4c$(x,y,1/4), Co $4b$(1/2,0,0), O1 $4c$(x,y,1/4), O2
$8d$(x,y,z)}
\begin{tabular*}{0.95\textwidth}{@{\extracolsep{\fill}} l|rrrrrrr}
\hline
T (K)  & 0.25      & 2         & 10        & 100       & 132       & 170       & 298         \\
\hline
a (\AA)& 5.2827(4) & 5.2823(5)  & 5.2824(5) & 5.2864(5) & 5.3041(6)  & 5.3241(4)  & 5.3405(6)   \\
b (\AA)& 5.3283(4) & 5.3274(5)  & 5.3275(5) & 5.3295(5) & 5.3349(6)  & 5.3401(4)  & 5.3513(6)   \\
c (\AA)& 7.4853(6) & 7.4846(8)  & 7.4847(8) & 7.4896(8) & 7.5051(9)  & 7.5274(6)  & 7.5471(8)   \\
\hline
x,Pr   & -0.0075(6)& -0.0090(10)& -0.0097(9)& -0.0066(9)& -0.0061(11)& -0.0050(10)& -0.0034(11) \\
y,Pr   & 0.0451(4) & 0.0431(6)  & 0.0441(6) & 0.0439(5) & 0.0410(6)  & 0.0377(5)  & 0.0324(2)   \\
x,O1   & 0.0794(5) & 0.0791(7)  & 0.0796(7) & 0.0781(7) & 0.0760(8)  & 0.0722(8)  & 0.0707(10)  \\
y,O1   & 0.4874(4) & 0.4861(6)  & 0.4865(6) & 0.4873(5) & 0.4884(6)  & 0.4902(5)  & 0.4928(16)  \\
x,O2   & -0.2917(3)& -0.2901(4) & -0.2902(4)& -0.2913(4)& -0.2883(5) & -0.2870(5) & -0.2848(14) \\
y,O2   & 0.2919(3) & 0.2912(4)  & 0.2904(4) & 0.2917(4) & 0.2896(5)  & 0.2874(4)  & 0.2848(14)  \\
z,O2   & 0.0399(2) & 0.0413(3)  & 0.0415(3) & 0.0401(3) & 0.0382(3)  & 0.0370(3)  & 0.0354(10)  \\
\hline
\end{tabular*}
\label{tab15}
\end{table*}

\begin{figure}
\includegraphics[width=\figwp\columnwidth,viewport=15 400 580 810,clip]{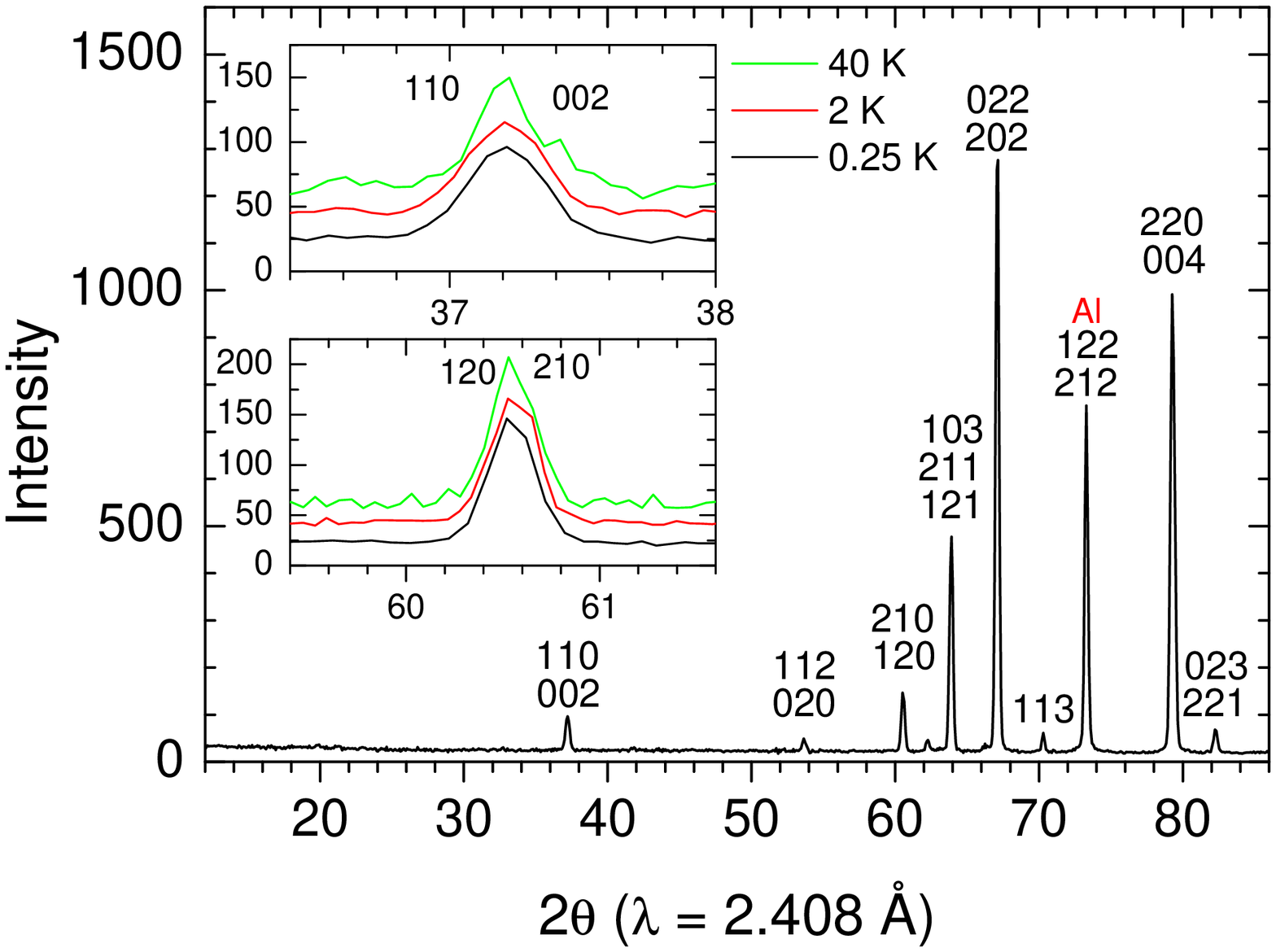}
\caption{Neutron diffraction of $y=0$ at low temperature.} \label{FigND00low}
\end{figure}

\begin{figure}
\includegraphics[width=\figwp\columnwidth,viewport=15 400 580 810,clip]{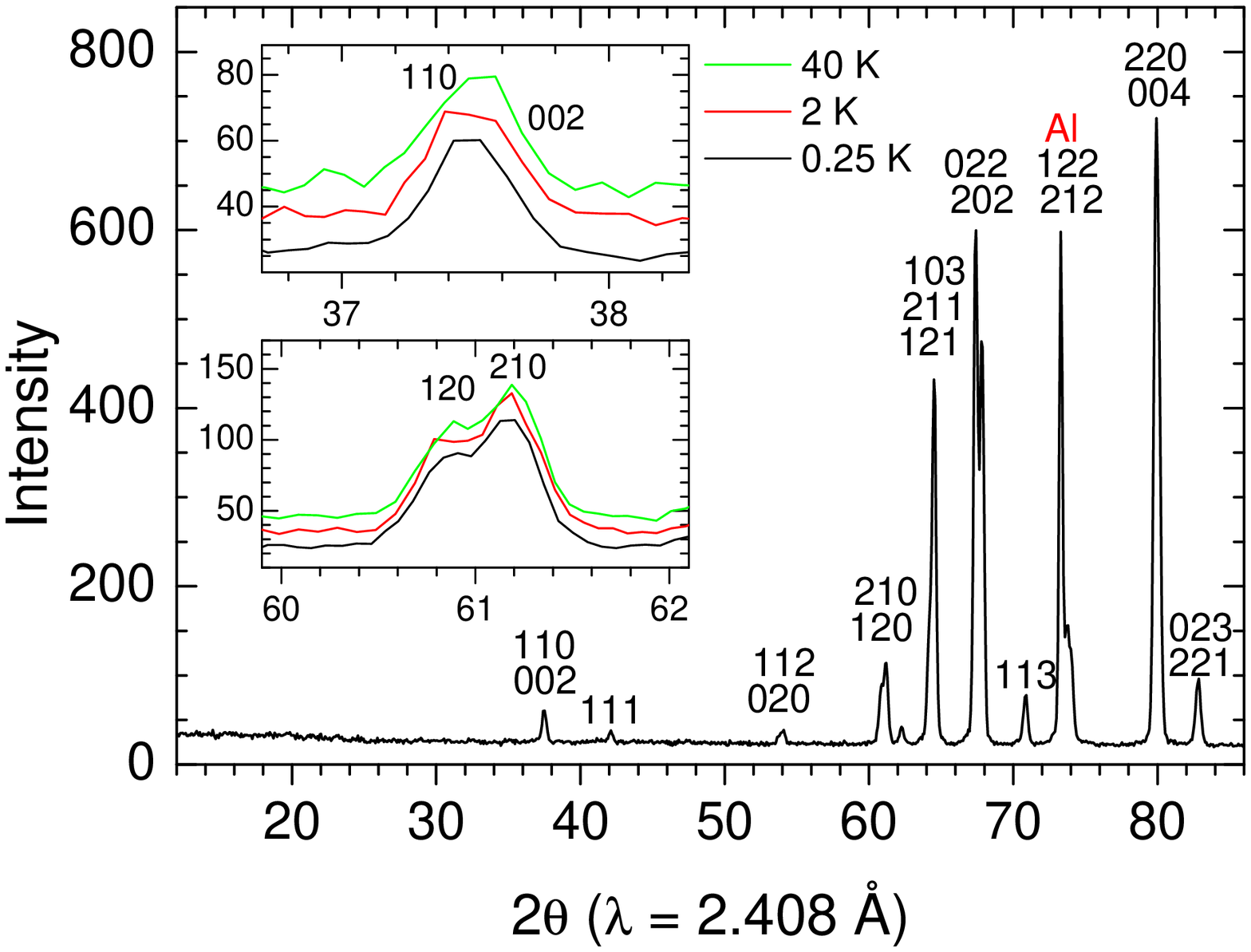}
\caption{Neutron diffraction of $y=0.15$ at low temperature.} \label{FigND15low}
\end{figure}

\begin{figure}
\includegraphics[width=\figwp\columnwidth,viewport=15 400 580 810,clip]{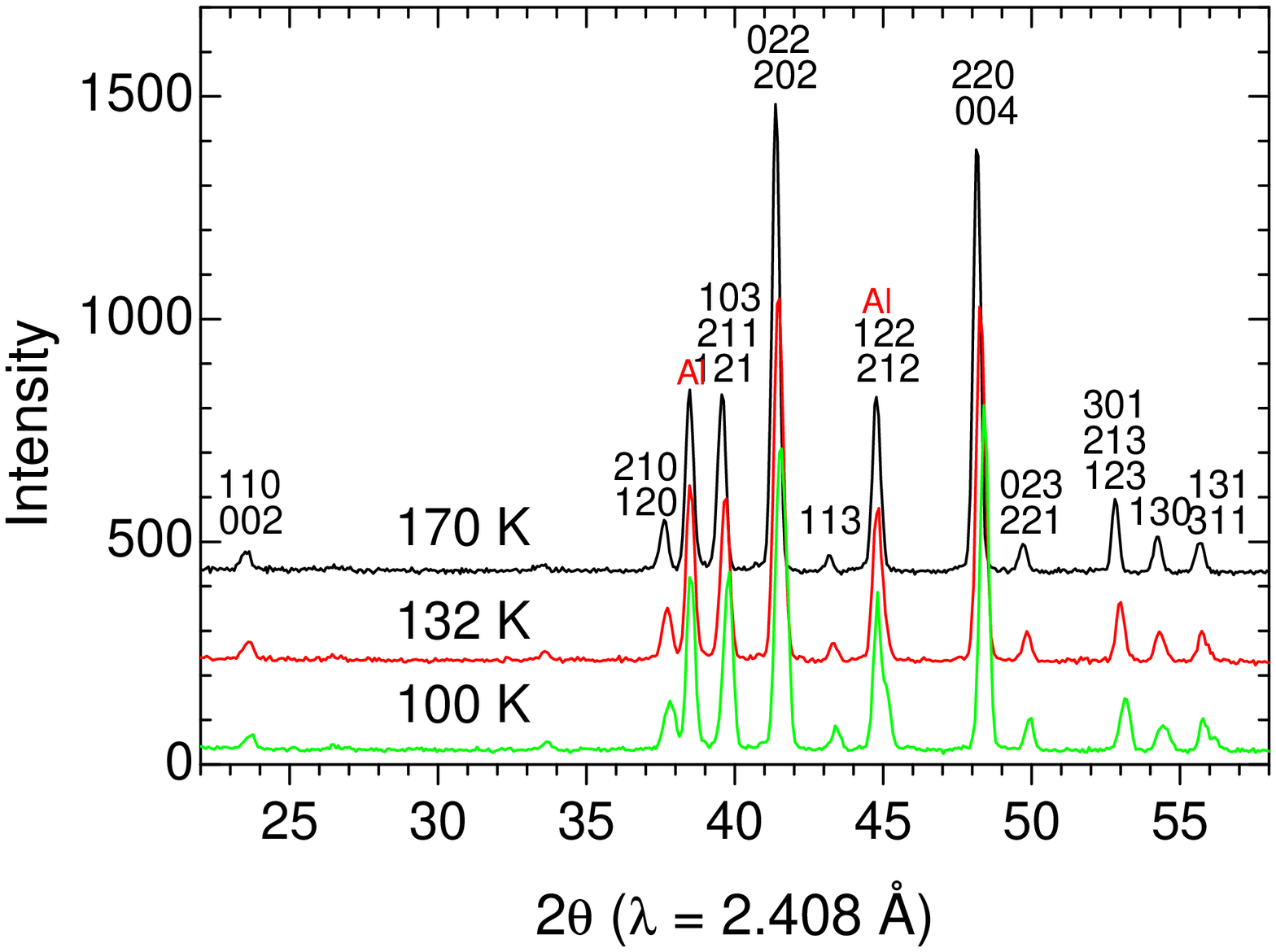}
\caption{Neutron diffraction of $y=0.15$ around the transition.} \label{FigND15high}
\end{figure}

\begin{figure}
\includegraphics[width=\figwp\columnwidth,viewport=0 100 500 827,clip]{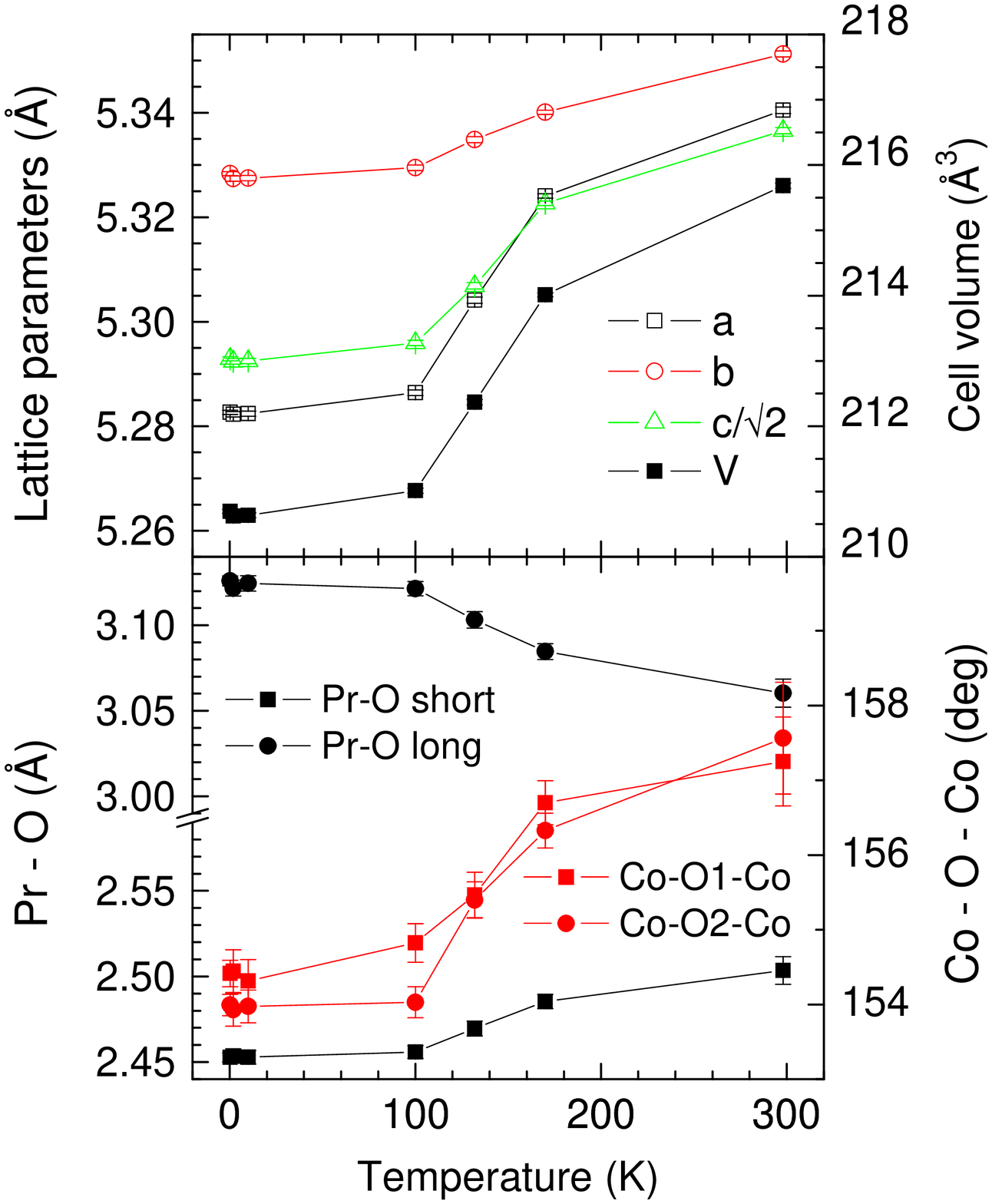}
\caption{Lattice parameters and bond lengths for $y=0.15$.} \label{Fig15Latt}
\end{figure}

The neutron diffraction pattern taken on \prycatri\ ($y=0$ and 0.15) are examplified together with
the FULLPROF fit in Fig.~\ref{FigND00low}, \ref{FigND15low} and \ref{FigND15high}. Although the
mixed-valence cobaltites are generally subjected to an oxygen deficiency, the occupation of oxygen
sites in present two samples is surely close to the ideal stoichiometry, the refined values being
$2.99 \pm0.01$ and $3.01\pm0.01$ per f.u., respectively. The values of other structural parameters
are summarized in Tables~\ref{tab00} and \ref{tab15}, and the unit cell volume and selected
interatomic distances and angles are plotted in Fig.~\ref{Fig15Latt}a,b. There is little change
with temperature for the $y=0$ sample, except common thermal expansion. On contrary, the $y=0.15$
sample shows a marked volume compression on cooling below $T_{MI}=132$~K, which is accompanied by
an increase of orthorhombic lattice distortion. Closer inspection reveals larger deviation of the
O-Co-O angles from ideal 180\st\ and some drop of the (Pr, Y, Ca)-O bonding distances, while the
Co-O distances remain practically unchanged. All these signatures are manifestations of the
decreased ionic size upon the Pr$^{3+}\rightarrow$Pr$^{4+}$ valence shift - see the similar
findings for \prcapet\ in Ref.~\cite{RefBaronGonzalez2010PRB81_054427}.

As the magnetic state is concerned, the $y=0$ sample shows below $T_C=55$~K a long range
ferromagnetic (FM) order of cobalt spins that is readily seen in magnetic measurements (see below)
but is only hardly visible in the neutron diffraction patterns as very weak increase of some
lines, mainly 110+002 and 200+112+020. The value of spontaneous moment determined from the neutron
data reaches $0.34\pm0.10 \mu_B$ at 0.25~K.

The detection of eventual FM ordering in $y=0.15$ is still more difficult since the cobalt
subsystem is now magnetically very dilute with only 0.12 LS Co$^{4+}$ ions per f.u. These moments
alone are below the detection limit of the neutron diffraction, but some chance is offered by very
low temperatures where full alignment of Pr$^{4+}$ moments (the $J'$=1/2 pseudospins) due to the
above-mentioned molecular field of $H_m \sim 17$~kOe could be anticipated and might add to the
cobalt FM order. Nonetheless, even at 0.25~K we failed to detect any observable diffraction
intensity that could prove an existence of long range ordering in the low-temperature phase of
$y=0.15$. Let us note that for prototypical \prcapet\ as well, no long-range FM order was found
\cite{RefBaronGonzalez2010PHYSPROC8_73} although the molecular field in that system is as high as
$H_m\sim75$~kOe \cite{RefHejtmanek2013EPJB86_305}. We address this issue below and suggest an
explanation.

\subsection{Magnetic measurements.}

\begin{figure}
\includegraphics[width=\figwp\columnwidth,viewport=0 20 545 827,clip]{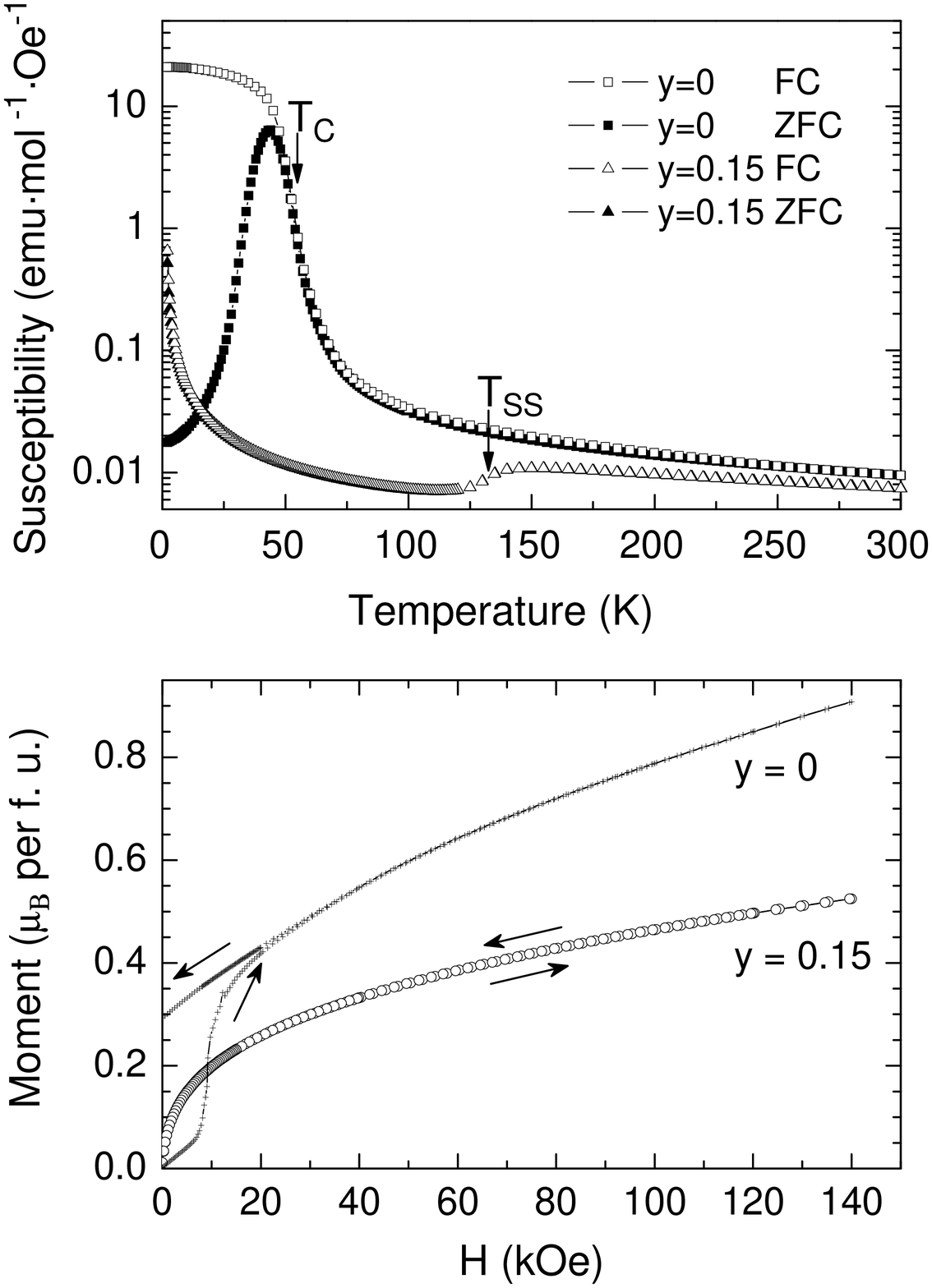}
\caption{Upper panel: The ZFC and FC susceptibility on \prycatri\ ($y=0$ and 0.15)
measured in DC field of 20~Oe. Note the logarithmic scale on the moment axis.
Lower panel: The magnetization in fields up to 140~kOe and back, taken at 2~K.} \label{FigZFCFC}
\end{figure}

The basic magnetic characterization of the \prycatri\ system was presented in our previous paper
\cite{RefHejtmanek2010PRB82_165107}. New data on ZFC/FC susceptibilities have been taken for low
DC fields. The results are given in upper panel of Fig.~\ref{FigZFCFC}, while the lower panel
illustrates the behavior in high magnetic fields, in particular it shows the magnetization curves
for the $y=0$ and 0.15 samples, taken at 2~K in the field range $0-140$~kOe and back.

The susceptibility data evidence a FM transition in the $y=0$ sample at $T_C=55$~K, determined by
inflection point on the fast rise of susceptibility curves, while the $y=0.15$ sample is
characterized by a drop of susceptibility that accompanies the $M-I$ transition at $T_{MI}=132$~K.
It should be noted that preformation of FM clusters at $250-270$~K, reported in some recent
studies of similar cobaltites \cite{RefElKhatib2010PRB82_100411,RefPhelan2013PRB88_075119}, is not
observed for present samples. In particular, neither ZFC/FC bifurcation nor finite remanence
extending to such high temperatures can be detected.

\begin{figure}
\includegraphics[width=\figwp\columnwidth,viewport=0 375 545 827,clip]{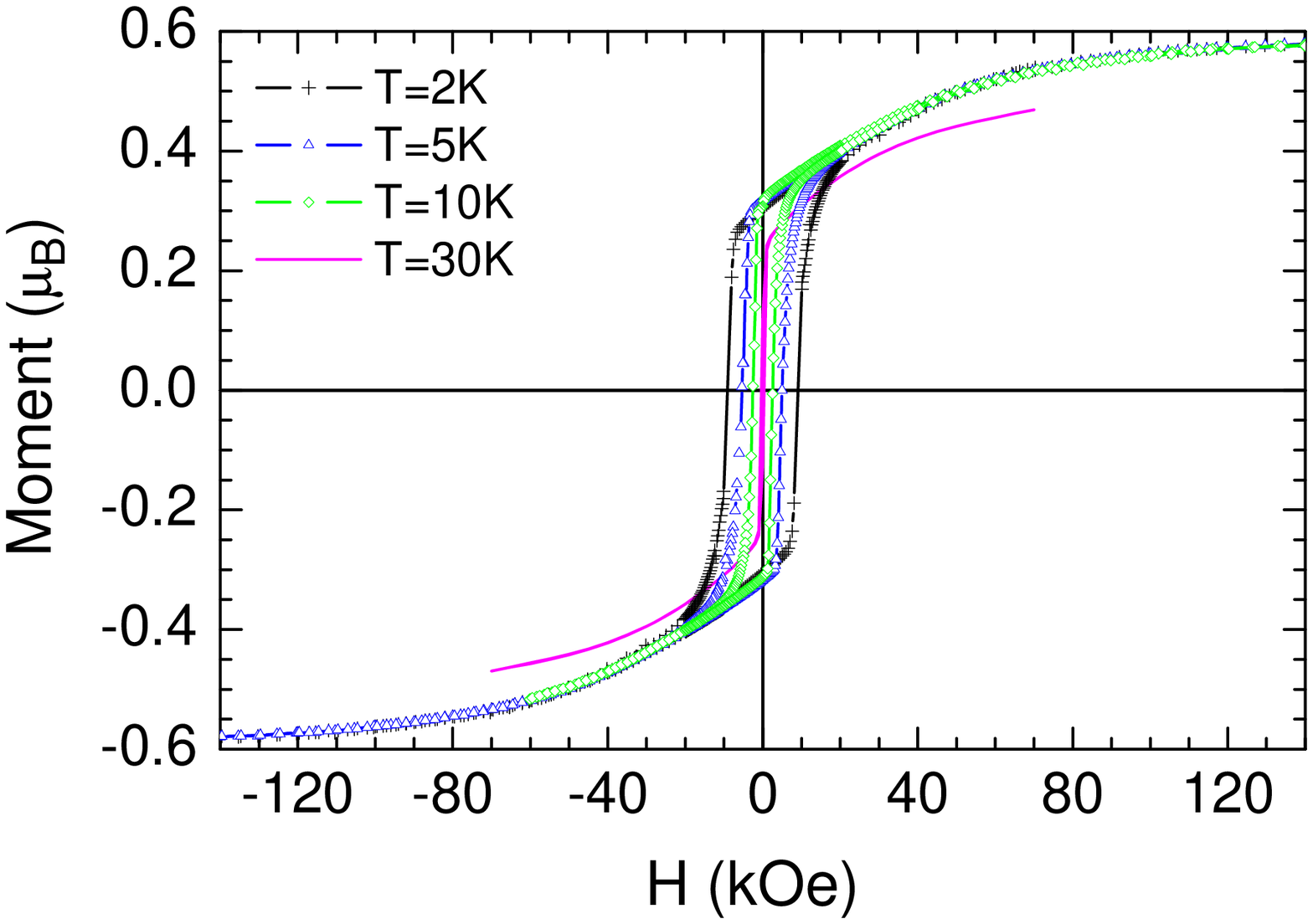}
\caption{The hysteresis loops for the \prcatri\ sample after subtraction of the Pr$^{3+}$ contribution.}
\label{FigMagn00}
\end{figure}

\begin{figure}
\includegraphics[width=\figwp\columnwidth,viewport=0 375 545 827,clip]{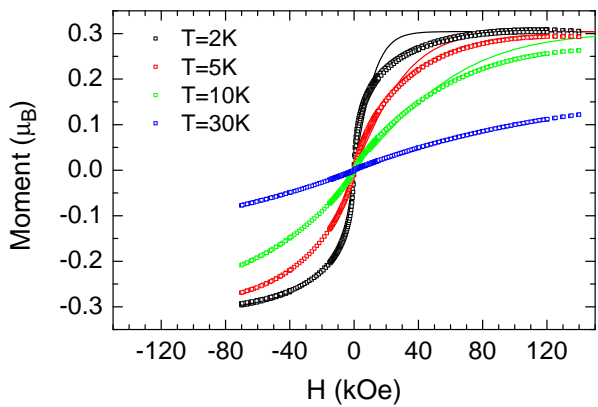}
\caption{The hysteresis loops for \prycatri\ ($y=0.15$). The theoretical Brillouin curves
for $S=1/2$ and $T=2$, 5 and 10~K are added for comparison, marked by the lines.}
\label{FigMagn15}
\end{figure}

The magnetization data for $y=0$ in lower panel of Fig.~\ref{FigZFCFC} show a significant
paraprocess that is superimposed on a hysteresis loop of relatively large coercivity (see
Fig.~\ref{FigMagn00}) that reach 9~kOe at 2~K. Similar paraprocess extending to very high fields
has been observed also for related manganites Pr$_{0.7}$Ca$_{0.3}$MnO$_3$
\cite{RefThomas1999JAP85_5384}. We thus ascribe this linear term to the Van Vleck susceptibility
of Pr$^{3+}$ and, taking experimental data on purely LS Co$^{3+}$ system PrCoO$_3$ in
polycrystalline form as a standard, $\chi_{vV}=0.0188$~\emumoloe, or equivalently
0.0034~$\mu_B$/kOe \cite{RefNovak2013JPCM}, we find for valence composition
Pr$^{3+}_{0.7}$Ca$^{2+}_{0.3}$Co$^{3+}_{0.7}$Co$^{4+}_{0.3}$O$^{2-}_{3}$ of the $y=0$ sample a
reduction of the paraprocess by factor 0.78. This is in a reasonable agreement with the actual
content of 0.70 Pr$^{3+}$.

After subtraction of the Pr$^{3+}$ related paraprocess, the magnetization of the cobalt subsystem
in $y=0$, presented in Fig.~\ref{FigMagn00}, saturates for 140~kOe at a value of 0.58~$\mu_B$/Co.
The spontaneous moment obtained from the extrapolation to zero field is, however, lower and
closely approaches the value 0.34~$\mu_B$, which was deduced from the neutron diffraction
refinement. In our interpretation, the ground state of \prcatri\ is based on so-called LS/LS phase
for Co$^{3+}$ and Co$^{4+}$ with theoretical FM moment of 0.30~$\mu_B$/Co. With increasing
external field, the IS/LS phase with theoretical FM moment of 1.70~$\mu_B$/Co seems to be
partially populated. For possibility of such two-phases coexistence see
Refs.\cite{RefSboychakov2009PRB80_024423,RefJirak2013JPCM25_216006}.

The magnetization data for the $y=0.15$ sample reveal a highly inhomogeneous phase that can be
characterized as cluster glass with complex behaviors both in the medium and high magnetic fields.
As shown in the lower panel of Fig.~\ref{FigZFCFC}, the high-field paraprocess is reduced compared
to \prcatri\ and corresponds now to 0.465 of the experimental value for PrCoO$_3$. This finding is
an independent confirmation of the Pr$^{3+}\rightarrow$Pr$^{4+}$ valence shift below $T_{MI}$. In
fact, it is in very good quantitative agreement with the valence composition, which was deduced
from Schottky peak analysis in Ref.\cite{RefHejtmanek2010PRB82_165107}~-~
Pr$^{3+}_{0.415}$Pr$^{4+}_{0.18}$Y$^{3+}_{0.105}$Ca$^{2+}_{0.3}$Co$^{3+}_{0.88}$Co$^{4+}_{0.12}$O$^{2-}_{3}$.

The magnetization of $y=0.15$ after the subtraction of the Pr$^{3+}$ paraprocess is presented in
Fig.~\ref{FigMagn15}. It saturates at 140~kOe on 0.30~$\mu_B$ per f.u., of which 0.12~$\mu_B$
should be attributed to the Co$^{4+}$ contribution in the LS/LS phase and remaining 0.18~$\mu_B$
is evidently the contribution of Pr$^{4+}$ pseudospins in the $y=0.15$ sample, supposing the value
of $g_{J'}$-factor close to 2. Let us note that such value is quite reasonable in view of our
theoretical estimates for Pr$^{4+}$, $g_x=3.757$, $g_y=0.935$ and $g_z=0.606$, which yield the
average value $\langle g_{J'} \rangle =2.07$ when a numerical integration over random orientation
of the crystallites is done.

\begin{figure}
\includegraphics[width=\figwp\columnwidth,viewport=0 375 545 827,clip]{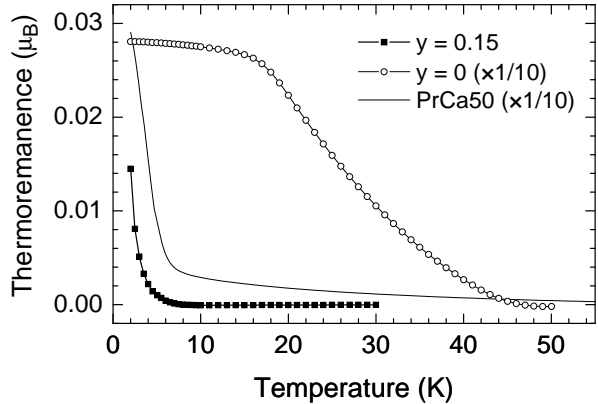}
\caption{The thermoremanent magnetization of \prycatri\ samples.
The data for $y=0.15$ are compared with the \prcapet\ sample possessing at least $20\times$
larger thermoremanence due to excessive defects.}
\label{Fig15rem}
\end{figure}

The magnetization curves observed for the $y=0.15$ sample at low temperatures deviate largely from
conventional FM behavior. At the first look, the $M vs. H$ dependence reminds a paramagnet close
to the saturation, but compared to standard Brillouin function the increase in low fields is
steeper, suggesting a presence of large spin entities with non-uniform distribution. We estimate
from the observed trend that superparamagnetic domains up to 100 cobalt ions are prevailing in the
$y=0.15$ sample. On the other hand, the increase in high fields is slower as if there were strong
AFM coupling  between up-grown FM regions or presence of surface states with large magnetic
anisotropy as known for so-called dead layer in nanoparticles. In closer inspection, one may also
notice certain opening of magnetization loops at the lowest temperatures, which is characterized
by coercitive field of about 200~Oe at 2~K. A finite but very weak remanent moment of
0.014~$\mu_B$/Co at 2~K quickly vanishes with increasing temperature (see Fig.~\ref{Fig15rem}).
Similar behavior has been observed also for other \prycatri\ samples with $M-I$ transition, as
well as for \prcapet\ where remanent moment at 2~K is, however, an order of magnitude larger
\cite{RefMarysko2012JAP111_07E110}.

\begin{figure}
\includegraphics[width=\figwp\columnwidth,viewport=0 330 545 827,clip]{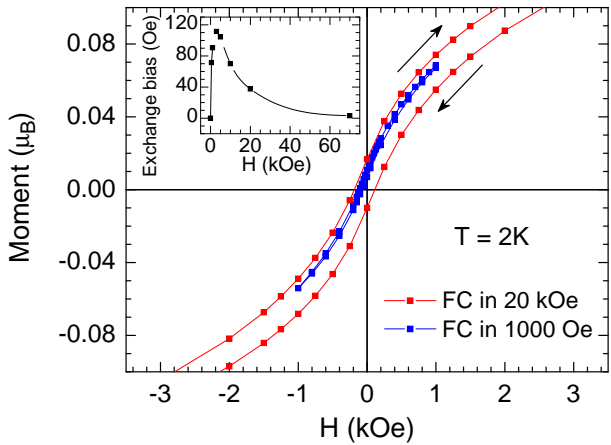}
\caption{The hysteresis loops for \prycatri\ ($y=0.15$), taken at 2~K after the cooling
in exchange bias fields 1 and 20~kOe.
The inset shows the exchange bias values (the abscissa shift of the loops)
in broader range of fields.}
\label{FigEB}
\end{figure}

Another signature of complex magnetic state of the $y=0.15$ sample is a presence of exchange bias,
demonstrated in Fig.~\ref{FigEB}. The effect is strongly dependent on the magnitude of bias field.

\begin{figure}
\includegraphics[width=\figwp\columnwidth,viewport=0 375 545 827,clip]{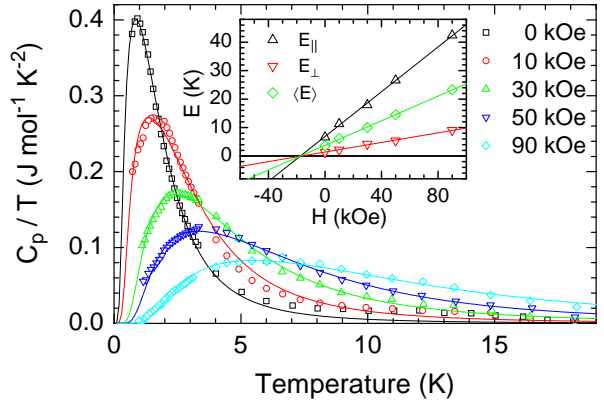}
\caption{The heat capacity divided by temperature for \prycatri\ ($y=0.15$), after
subtraction of lattice and nuclear terms. The full lines present the theoretical fit based on the
broadening due to anisotropic Zeeman splitting, supposing the axial symmetry of the $g$-factor.
The actual values of Zeeman energy are plotted in the inset.}
\label{FigCpFit}
\end{figure}

\section{Discussion}

The cobaltites are systems with large complexity. In order to understand the physical properties
of the mixed-valence cobaltites of 30\% doping range, at least in an illustrative way, the IS/LS
scenario for Co$^{3+}$/Co$^{4+}$ is generally applied. It is worth mentioning that this scenario
implies carriers in two bands of very different characters and mobilities. The $e_g$
quasiparticles are light but move in a disordered background of nearly localized $t_{2g}$ charges
and spins, and become strongly scattered, while the $t_{2g}$ quasiparticles though heavy and
short-living see a practically uniform background because of very fast fluctuations of the $e_g$
electron subsystem. The decreased scattering in the latter case may cause that the overall
conductivity is likely contributed by both the $e_g$ and $t_{2g}$ channels
\cite{RefAugustinsky2013PRL110_267204}. With decreasing temperature, the system \lasrtri\ of broad
$e_g$ band maintains IS/LS phase, whereas the narrow-band systems like \prcatri\ or \ndcatri\
exhibit a gradual crossover towards lower spin states. In contrast to true metallic conductivity
of \lasrtri\, the electrical resistivity observed on \prcatri\ or \ndcatri\ shows a steady
increase with decreasing temperature but extrapolates to a finite value instead of diverging. This
suggests that also in this latter case the LS/LS phase of 30\% $t_{2g}$ hole doping should be
considered as instrinsically metallic \cite{RefJirak2013JPCM25_216006}.

The spin state crossover in the yttrium containg samples \prycatri\ is much more abrupt since it
is accelerated by the electron transfer from Pr$^{3+}$ to Co$^{4+}$ at the $M-I$ transition. In
the LS/LS ground state, the carrier concentation in the Co-O subsystem is thus reduced, in
particular for the $y=0.15$ sample down to 12\% $t_{2g}$ hole doping. Following the observed
resistivity dependence down to 6~K, the actual regime of conduction is variable range hopping
\cite{RefHejtmanek2010PRB82_165107,RefHejtmanek2013EPJB86_305}. This means that the charge
transport is realized via localized states close to Fermi level or, in other words, the doping
level in the $y=0.15$ sample is below the mobility edge for the $t_{2g}$ band of LS/LS phase.

The difference between the $y=0$ and 0.15 samples is manifested also in their magnetic state. The
$y=0$ sample behaves at the lowest temperatures as a standard FM system with broad hysteresis
loops that evidence an existence of domain structure. The long range magnetic order and
spontaneous moments of about 0.30~$\mu_B$/Co are found by both the neutron diffraction and
magnetization measurements. We may mention, however, that the transition from the paramagnetic
state is unconventional. First, no heat capacity anomaly is observed at $T_C=55$~K. Another
feature reported in some previous studies of \prcatri\ is the observation of nanosize FM objects
at temperatures as high as $\sim 250$~K. These have been detected by both the small angle neutron
scattering and ZFC/FC bifurcation of the susceptibility \cite{RefElKhatib2010PRB82_100411}. The
lack of the latter anomaly in present study suggests that if such preformed FM objects are
intrinsic to \prcatri, their size and number of are likely sample dependent.

The long-range FM order is absent in our $y=0.15$ sample as well as in \prcapet, which is known as
prototype of the $M-I$ transition in Pr-based cobaltites. Three magnetic contribution are
effective in the insulating phase of these systems~-~spins of LS Co$^{4+}$, pseudospins of the
Pr$^{4+}$ ground doublet and singlet states of Pr$^{3+}$ with extremely large Van Vleck
susceptibility $\chi_{vV}$=0.0188~\emumoloe, or equivalently 0.0034~$\mu_B$/kOe ($\chi_{vV}$ for
the spin and orbital singlet LS Co$^{3+}$ is two orders of magnitude smaller, 0.0002~\emumoloe
\cite{RefGriffith1957TFS53_601,RefKamimura1966JPSJ21_484}, and has thus negligible effect on the
total magnetization).

It has been shown above that the paraprocess due to Pr$^{3+}$ contribution is simply separable,
and consistent data with respect to the Pr$^{3+}\rightarrow$Pr$^{4+}$ valence shift are obtained.
The resolution of the LS Co$^{4+}$ and Pr$^{4+}$ contributions appeared impossible, but their
overall magnetization in Fig.~\ref{FigMagn15} is suggesting for a magnetically non-uniform phase
that can be classified as the cluster glass or glassy ferromagnetism \cite{RefWu2003PRB67_174408}.
We may conclude, based on the initial slope of magnetization curves and presence of finite
thermoremanence in the $y=0.15$ sample, that the prevailing superparamagnetic domains with easy
saturation in low fields (average size of about 100 atoms) are coexisting with minor FM regions.
In addition, there is surely certain population of smaller magnetic clusters or even individual
spins that are manifested by the frequency dependent AC susceptibility below 10~K (not shown
here).

The pending question is the origin of relatively strong molecular field that acts on Pr$^{4+}$
pseudospins in the systems with occurrence of the $M-I$ transition, in spite of the highly diluted
and therefore inhomogeneously distributed Co$^{4+}$ spins. To illustrate the problem, we revisit
here the low-temperature Schottky data measured on the polycrystalline sample $y=0.15$ in our
previous paper \cite{RefHejtmanek2010PRB82_165107}. The Schottky peaks are broadened with respect
to the ideal Schottky form, but as seen in Fig.~\ref{FigCpFit} an excellent fit can be obtained
for $g$-factor of axial anisotropy, $g_\parallel=5.86$, $g_\perp=1.25$, irrespective the strength
of applied field. Similar quasi-axial anisotropy of $g$-factor has been found in recent analysis
of the Pr$^{4+}$-related Schottky peaks in \prcapet\ \cite{RefHejtmanek2013EPJB86_305}. Important
implications can be drawn from the analysis. First, the ratio $g_\parallel$/$g_\perp \sim=4.7$
agrees very well with the pseudoaxial theoretical ratio for Pr$^{4+}$, 2$g_x$/($g_y$+$g_z$)=4.9,
suggesting that there is little or no additional inhomogeneous broadening. Hence, the Pr$^{4+}$
pseudospins experience a molecular field of uniform strength, $H_m=17$~kOe, although they are
distributed randomly in the magnetically non-uniform phase. The origin of the molecular field
should be, therefore, more complicated than in the case of Nd$^{3+}$ containing perovskites with
FM ground state. Second, the Schottky peak profile retains the same $g$-anisotropy related
broadening even at zero field, which means that the molecular field present spontaneously in the
single crystal grains is not oriented along a particular crystallographic axis but behaves as in a
random anisotropy system. Third, the shift of Schottky peaks with increasing external field is
strictly linear suggesting an extreme softness, which is again in variance with behavior in FM
systems with domain structure. It is also of interest that the absolute $g$-values deduced from
the shift in applied field, in particular the average value $\langle g_{J'} \rangle =3.22$ is much
larger larger than the theory and also the saturated magnetization give ($\langle g_{J'} \rangle
\sim 2$). This discrepancy suggests that there is a spin polarizable medium in the Pr$^{4+}$
surroundings that acts as an enhancement of external field. This issue has been largely discussed
in relation to an opposite effect in Nd-based cobaltites \cite{RefJirak2013JPCM25_216006}.

\section{Conclusions}

The neutron diffraction and magnetic study have been performed on two cobaltite systems \prycatri\
($y=0$ and 0.15) with practically ideal oxygen stoichiometry. Both samples are presumably in a
dynamic mixture of IS Co$^{3+}$/LS Co$^{4+}$ states at room temperature and exhibit with
decreasing temperature a spin-state crossover. In both cases, the ground state is identified,
based on observed magnitudes of magnetic moments and robustness up to high fields, with a mixture
of LS Co$^{3+}$/LS Co$^{4+}$ states.

The \prcatri\ sample with formal concentration of 0.30 LS Co$^{4+}$ in the diamagnetic background
of LS Co$^{3+}$ shows characteristics of common FM phase ($T_C=55$~K) with prevailing long-range
order. Its electronic properties evidence the intrinsic metallicity despite the granularity of
ceramic sample.

The behaviour of yttrium substituted samples is much more complex because of a partial
Pr$^{3+}\rightarrow$Pr$^{4+}$ and Co$^{4+}\rightarrow$Co$^{3+}$ valence shift, which diminishes
the carrier doping in the Co-O subsystem and facilitates also the spin-state crossover to the
LS/LS phase. This transition, accompanied with a drop of electrical conductivity, occurs for
$y=0.15$ at $T_{MI}=132$~K. The transport data published earlier show that an effective conduction
mechanism at the low-temperature phase is the variable range hopping, which anticipates a
tunneling of $t_{2g}$ carriers between more distant Co sites close in energy. As the magnetism is
concerned, the present neutron diffraction, performed down to 0.25~K, does not show any long-range
ordering of Co$^{4+}$ and/or Pr$^{4+}$ moments, and also the magnetization loops display no or
negligible opening and are characteristic rather for a glassy state. The main finding is the
observation of uniform internal field acting on Pr$^{4+}$ pseudospins, which is surprising for the
low-temperature phase of $y=0$ with reduced concentration of cobalt spins (formally 0.12 LS
Co$^{4+}$ per f.u.).

The issue deserving most attention is the origin of FM interactions that are responsible for the
long-range ordering in the 30\% hole doped \prcatri\ or \ndcatri, as well as for the strong
molecular field acting on rare earth ions in the \prycatri\ systems with severely reduced hole
dopings. Whatever the character of LS Co$^{3+}$/LS Co$^{4+}$ mixture, either localized or forming
a very narrow $t_{2g}$ band, the FM interactions in a pure phase are unlike. Namely, the
superexchange interactions between LS Co$^{4+}$ (the $t_{2g}^5$ configuration) should be of AFM
type according to Goodenough-Kanamori rules, and also when collective $t_{2g}$ states and
single-band Hubbard model are considered, no spontaneous FM ordering can be foreseen. The double
exchange interactions that are effective in the room temperature IS/LS phase are neither possible.
In our opinion, a possible explanation stems from magnetic defects that arise due to minor oxygen
vacancies inherently present in doped cobaltites. This results in a reduced coordination of nearby
cobalt ions, likely the pyramidal one, which stabilizes local HS Co$^{3+}$ states. Their
concentration is estimated based on the near oxygen stoichiometry in our samples to be $\sim2$\%
or less. We suggest that these very dilute defects, clearly below the percolation limit, interact
ferromagnetically via itinerant $t_{2g}$ holes. The spin polarization of the narrow $t_{2g}$ band
then mediates the nearly homogeneous molecular field over the crystal lattice.

To summarize: The sole existence of FM order in the LS/LS cobaltite phases is a general problem.
Our explanation  relies on a scenario of magnetic defects, which represent a non-homogeneity, but
polarize $t_{2g}$ carriers at the top of oxygen $\pi$ hybridized band. Irrespective the nature of
these defects and their actual location, the spin density is thus distributed over the whole
sample.

\textbf{Acknowledgments}. We thank J. Kune\v{s} and C. Leighton for stimulating comments. This work was
supported by Project No.~204/11/0713 of the Grant Agency of the Czech Republic.


\end{document}